\documentclass[prd,aps,a4paper,nofootinbib,twocolumn]{revtex4}  %-1

\newif\ifusesec
\usesectrue  
   
\usepackage{graphicx} 
\usepackage{mathrsfs}
\usepackage{amsmath,amsfonts,amssymb}
\usepackage{multirow}

\newcommand{\beq}{\begin{equation}}
\newcommand{\eeq}{\end{equation}}
\newcommand{\bea}{\begin{eqnarray}}
\newcommand{\eea}{\end{eqnarray}}
\begin{document}

\title{High post-Minkowskian gravitational waveform for hyperbolic encounters in the extreme-mass-ratio limit}

\author{Andrea Geralico}
\affiliation{
Istituto per le Applicazioni del Calcolo ``M. Picone,'' CNR, I-00185 Rome, Italy
}

\date{\today}

\begin{abstract}
The frequency-domain waveform emitted by a two-body scattering process is computed in the extreme-mass-ratio limit through the fifth post-Minkowskian (PM) order (i.e., $O(G^5)$) and the fractional sixth post-Newtonian (PN) order.
The current accuracy of the scattering waveform obtained by quantum amplitude methods is the one-loop level corresponding to the 3PM order, whereas the 4PM waveform is known up to the 2PN order only as derived within the traditional multipolar-post-Minkowskian formalism.
Direct comparison between these waveforms to the first order in the mass ratio shows that they differ at most by the effect of an angular-independent time shift, leading to a complete physical agreement at the same level of accuracy.
The new results at the 4PM and 5PM orders thus provide a benchmark for future multiloop calculations.
The waveform is also used to obtain the radiated energy at $O(G^6)$, extending it from 3PN to 6PN level.
\end{abstract}

%\pacs{04.20.Cv, 98.58.Fd}

\maketitle

{\it Introduction.}--
The events observed so far by current gravitational wave detectors are associated with the coalescence of compact binaries, whereas bremsstrahlung events still appear to be out of reach \cite{Henshaw:2025arb}.
However, next-generation ground-based detectors as well as space-based observatories are planned to be soon operational, with a significantly increased sensitivity allowing for the first observations of close hyperbolic encounters \cite{Purrer:2019jcp,Dandapat:2023zzn,Bini:2023gaj}.
The latter are expected to occur, e.g., in regions dense with stellar remnants, like galactic nuclei and globular clusters, hosting a large number of black holes and neutron stars, with a peculiar emission of gravitational wave bursts during the rapid interaction phase in which the two bodies either scatter and fly past each other or become bound  \cite{Kocsis:2006hq,Berry:2012im,Mukherjee:2020hnm}.  

The precision demands of future experiments require high-precision waveform models, which can be built up by improving the analytical description of compact binary dynamics.
Most of the effort over the years has focused on bound orbits in a post-Newtonian (PN) framework (see, e.g., Ref. \cite{Blanchet:2013haa}), by using analytical perturbation methods, including the traditional Multipolar-Post-Minkowskian (MPM) formalism \cite{Blanchet:1985sp,Blanchet:1986dk,Blanchet:1987wq,Blanchet:1989ki,Damour:1990gj,Damour:1990ji,Blanchet:1992br}, the gravitational self-force (GSF) approximation \cite{Barack:2018yvs}, and the effective field theory (EFT) approach \cite{Goldberger:2004jt,Porto:2016pyg,Levi:2018nxp,Foffa:2013qca,Foffa:2021pkg}, in combination with numerical-relativity (NR) simulations.
All that information is then collected together within the effective-one-body (EOB) formalism \cite{Buonanno:1998gg,Damour:2000we}, which is able to produce highly accurate inspiral-merger-ringdown waveforms currently adopted in gravitational wave searches.

In the case of hyperbolic encounters the velocities can reach high values, so that the PM approximation is more suited, since it is valid for arbitrary velocities.
The leading-PM-order contribution to the waveform was computed long ago in the time-domain for nonspinning bodies by Kovacs and Thorne \cite{Kovacs:1977uw,Kovacs:1978eu} in the framework of classical perturbation theory (see also Refs. \cite{Turner:1977,Turner:1978zz} for PN-based results).
This seminal result was recently revisited by using quantum-field-theory-based techniques developed in high energy physics and EFT \cite{Jakobsen:2021smu,Mougiakakos:2021ckm,DiVecchia:2021bdo}.
The classical waveform can be directly computed from scattering amplitudes \cite{Cristofoli:2021vyo}. 
At the lowest order it is given by an integral of a five-point tree-level amplitude with one graviton emission.
The next order (one-loop) contribution to the waveform was recently computed independently by different groups \cite{Brandhuber:2023hhy,Herderschee:2023fxh,Elkhidir:2023dco,Georgoudis:2023lgf}.
This is the state-of-the-art of amplitude-based computations, i.e., 3PM order.
Going beyond this level requires a significant build-up of technology and the development of advanced multiloop integration techniques in order to efficiently evaluate master integrals. 

The 4PM (two-loop) order waveform is currently known at the fractional 2PN accuracy. 
It has been recently computed in the frequency domain in Ref. \cite{Bini:2024ijq} by using the MPM formalism, which expresses the waveform as a sum over radiative multipolar contributions combined with a double PM-PN power series in $G$ and $\frac1c$.
In order to obtain highly accurate waveforms the MPM formalism needs to know the radiative multipole moments and their relations with the source multipole moments up to a high PN order as well as all nonlinear couplings occurring at each PM order leading to hereditary integrals over the past history of the source, including tails, tails of tails, memory, and higher nonlinear interactions \cite{Blanchet:1992br}.
In addition, when computing the frequency-domain waveform one must take into account also the effect of radiation-reaction on the scattering motion, starting at the 2.5PN order. 
The (leading) mass-type quadrupole moment is currently known at the 4PN level \cite{Blanchet:2023bwj,Blanchet:2023sbv}, whereas the radiation-reaction corrected dynamics at the 3.5PN level \cite{Bini:2025rng}.

The aim of the present paper is the computation of the frequency-domain scattering waveform within the first-order self-force (1SF) approximation. The latter approach in general holds for any value of the velocity and can access the strong field regime, but only in the extreme-mass-ratio limit, i.e., to first order in the binary's mass ratio.
However, the analysis done here uses a combined PM-PN approximation scheme, i.e., at each PM order all quantities are expanded for small values of the velocity up to a high PN level.
I use the standard Teukolsky formalism complemented by the Mano, Suzuki and Takasugi (MST) method to construct the waveform modes from the Weyl scalar $\psi_4$, which encodes all the information on the radiation emitted by the system (see, e.g., Ref. \cite{Sasaki:2003xr} for a review).
The analytical results are expressed as double PM-PN expansions, and are accurate to the 5PM-6PN level, i.e., two PM orders beyond the amplitude-based results.
As a byproduct I also compute the 1SF radiated energy at the 6PM order with the same PN accuracy.

Units are chosen so that $G=1=c$, unless otherwise specified.

{\it Gravitational waveform for hyperboliclike orbits.}--
The classical waveform $h_{\mu \nu} = g_{\mu \nu}-\eta_{\mu \nu}$ has the PM expansion 
\bea 
\label{hmunu}
h_{\mu \nu}&=&G h^{\rm 1PM}_{\mu \nu}+ G^2 h^{\rm 2PM \, or\, tree}_{\mu \nu} + G^3 h^{\rm 3PM\, or\,  one-loop}_{\mu \nu} \nonumber\\
&+&  G^4 h^{\rm 4PM \, or\,  two-loop}_{\mu \nu} 
+G^5 h^{\rm 5PM \, or\,  three-loop}_{\mu \nu}\nonumber\\
&+&O(G^6)\,.
\eea
The linear contribution $G h^{\rm 1PM}_{\mu \nu}$ is purely stationary in the time domain, so that it is localized at zero frequency ($\omega=0$) in the frequency domain.
It is useful to introduce a reduced complex asymptotic waveform
\beq
W=\frac{c^4}{4 G} \lim_{r\to \infty}(r( h_+ -  i h_\times))\,,
\eeq
where $h_+$ and $h_\times$ denote the two transverse-traceless components, and the prefactor $\frac{c^4}{4 G}$ is used to simplify the corresponding PN expansion, in such a way that the leading PN contribution to $W$ is just the classic quadrupole formula, and its non-zero frequency part ($\omega\not=0$) starts at $O(G^1)$, i.e., 
\beq
W\underset{\omega\not=0}=GW^{G^1}+G^2W^{G^2}+G^3W^{G^3}+G^4W^{G^4}+O(G^5)\,.
\eeq
[See Table \ref{table1} for the classical versus quantum nomenclature.]
Therefore, the combined PM-PN expansion of the waveform is understood to be carried out in terms of fractional corrections to the leading-order quadrupole term.

% table 1

\begin{table}  
\caption{\label{table1} Dictionary between the quantum nomenclature and the classical PM one. 
}
\begin{ruledtabular}
\begin{tabular}{lll}
tree & $h^{\rm 2PM}_{\mu \nu}$ & $W^{G^1}$\\ 
one-loop & $h^{\rm 3PM}_{\mu \nu}$ & $W^{G^2}$\\
two-loop & $h^{\rm 4PM}_{\mu \nu}$ & $W^{G^3}$\\
three-loop & $h^{\rm 5PM}_{\mu \nu}$ & $W^{G^4}$\\
\end{tabular}
\end{ruledtabular}
\end{table}

The classically-relevant part of the 5-point momentum-space scattering amplitude ${\mathcal M}(\epsilon, k, q_1, q_2) $, where $\epsilon$ is a null polarization vector, $k$ is the null momentum of the emitted graviton, and $q_1=p_1-p'_1, q_2=p_2-p'_2$ are the transferred momenta, has been derived in Refs. \cite{Brandhuber:2023hhy,Herderschee:2023fxh,Elkhidir:2023dco,Georgoudis:2023lgf}.
Its {\it integrated} version ${\mathcal M}^{\rm int}(k)$ is related via the formalism of Ref. \cite{Kosower:2018adc} to the frequency-domain classical (reduced) waveform by $W(\omega, \theta,\phi)=\frac{2i}{\kappa}{\mathcal M}^{\rm int}(k)$, where $\kappa \equiv \sqrt{32 \pi G}$, and $\omega, \theta,\phi$ parametrize $k$. 
The comparison of the EFT waveform at the one-loop level with the MPM waveform requires addressing delicate issues, including how to fix the spatial direction of the various vectors defining the waveform in the center-of-mass frame within which the MPM formalism has been developed, as discussed in Refs. \cite{Bini:2023fiz,Bini:2024rsy}. I will follow definitions and conventions used there to relate the 1SF waveform obtained below with the EFT and MPM waveforms in the extreme-mass-ratio limit.
The PM expansion corresponds to a power series expansion in the inverse dimensionless angular momentum $\frac1j= \frac{G m_1 m_2}{cJ}$, or equivalently in powers of $GM/b$, where $b$ denotes the impact parameter, and $M=m_1+m_2$ is the total mass of the system (with $m_2>m_1$). 
The impact parameter $b$ is related to $j$ by $\frac1j= \frac{G M h}{ b p_\infty}$, where $h=E/Mc^2\equiv\sqrt{1+2\nu(\gamma-1)}$ is the dimensionless incoming center-of-mass energy, $\nu=m_1m_2/M^2$ the symmetric mass ratio, and $p_\infty=\sqrt{\gamma^2-1}$ a velocity parameter controlling the PN expansion.

According to the GSF approach the Einstein field equations are solved by using an expansion in powers of the small mass ratio $q=\frac{m_1}{m_2}\ll1$, so that $h_{\mu \nu}=O(q)$ to first order.
Recalling the definition of the symmetric mass ratio, one can then consistently expand all quantities in powers of $\nu=q+O(q^2)$ at fixed $M$.
Let the smaller mass be moving along a hyperboliclike geodesic orbit on the equatorial plane ($\theta=\frac{\pi}{2}$) of a Schwarzschild spacetime with parametric equations $x^\mu =x_p^{\mu}(\tau)$, $\tau$ denoting the proper time.
To first order in $\nu$ the waveform can be computed in the framework of first-order perturbation theory by using the Teukolsky formalism. I refer to Ref. \cite{Sasaki:2003xr} for definitions, notation and conventions.
 
The Weyl scalar $\psi_4$ is asymptotically related to the two independent polarizations $h_+$ and $h_\times$ of the gravitational waves by
\beq
\label{hdef}
\psi_4(r\to\infty)\sim-\frac12\ddot h\,,
\eeq
where $h\equiv h_+-ih_\times$, a dot denoting time derivative.
$\psi_4$ satisfies the Teukolsky equation with spin-weight $s=-2$, and can be decomposed as 
\beq
\label{sep}
\psi_4= \frac1{r^4}\int\frac{d\omega}{2\pi}e^{-i\omega t}\sum_{lm}\,\,R_{lm\omega}(r)\,\, {}_{-2}Y_{lm}(\theta,\phi)\,,
\eeq
where ${}_{s}Y_{lm}(\theta,\phi)$ are spin-weighted spherical harmonics (SWSH).
The radial function $R_{lm\omega}(r)$ satisfies the inhomogeneous Teukolsky equation with source term $T_{lm\omega}(r)$. 
The asymptotic solution representing purely outgoing waves is given by 
\beq
R_{lm\omega }(r\to\infty)\sim Z^\infty_{lm\omega} r^3e^{i\omega r_*}\,,
\eeq
where $r_*$ is the tortoise coordinate, and $Z^\infty_{lm\omega}$ is the amplitude (see Eqs. (20) and (30) of Ref. \cite{Sasaki:2003xr}).
The latter is constructed through the energy momentum tensor associated with the source, which has the form of a Dirac-delta function with support on the worldline of the perturbing body (see Eq. (21) of Ref. \cite{Sasaki:2003xr}). As a result, the amplitude will be proportional to its mass $m_1$, and the trajectory can be taken as a geodesic of the background field to first order in the mass ratio.   

The asymptotic form of $\psi_4$ implies
\beq
\label{h}
h=\frac{4G}{r}\sum_{lm}\int\frac{d\omega}{2\pi}{\mathcal W}_{lm}(\omega)e^{-i\omega(t-r_*)}\,\,{}_{-2}Y_{lm}(\theta,\phi)\,,
\eeq
where 
\beq
\label{Wlmdef}
{\mathcal W}_{lm}(\omega)\equiv\frac{Z^\infty_{lm\omega}}{2\omega^2}\,.
\eeq
The frequency-domain rescaled waveform ${\mathcal W}=rh/4G$ can then be written as 
\beq
\label{Wfin}
{\mathcal W}(\omega,\theta,\phi)=\sum_{lm}{\mathcal W}_{lm}(\omega)\,{}_{-2}Y_{lm}(\theta,\phi)
\,.
\eeq
The waveform modes \eqref{Wlmdef} are integrals of the type
\beq
\label{Wlmdef2}
{\mathcal W}_{lm}(\omega)=\int dt e^{i(\omega t-m\phi_p(t))}{\mathcal F}_{lm\omega}(r_p(t))\,,
\eeq
with the function ${\mathcal F}_{lm\omega}(r_p(t))$ evaluated at the particle position $r=r_p(t)$, and are computed by using the MST solutions satisfying the retarded boundary conditions of ingoing radiation at the horizon and upgoing at infinity \cite{Mano:1996mf,Mano:1996vt}.
Each value of $l\geq2$ contributes half a PN order, so that one needs to sum terms up to $l=14$ to reach the fractional 6PN accuracy.

In order to compute the integrals \eqref{Wlmdef2} it is convenient to parametrize the geodesics in a quasi-Keplerian form, 
taking then both the PM and PN expansions of the integrand in powers of $G$ and $\frac1c$, respectively, leading to Bessel and iterated-Bessel integrals \cite{Bini:2021jmj,Bini:2021qvf,Bini:2023fiz}.

{\it Structure of Fourier integrals.}--
At the lowest PM orders $G^1$ and $G^2$ the SWSH components of the reduced waveform, $W^{G^1}$ and $W^{G^2}$, are expressed as linear combinations of  modified Bessel functions of order $0$ and $1$, $K_0(u)$ and $K_1(u)$, and of the exponential function $e^{-u}$ (see, e.g., Ref. \cite{Bini:2023fiz}), with argument given by the following dimensionless version of the frequency (assuming $\omega>0$)
\beq
u \equiv \frac{\omega b}{p_\infty}\,,
\eeq
and coefficients which are polynomials in $u$ with degree increasing with the PN order.
In fact, the Fourier integrals \eqref{Wlmdef2} all are of the form
\beq 
\label{Q}
Q_{\alpha}(u) \equiv \int dT \frac{e^{iu T}}{(1+T^2)^{\alpha}} \,,
\eeq
which can be expressed for a generic index $\alpha$ in terms of modified Bessel $K$ functions
\beq
\label{Q_integ}
Q_{\alpha}(u) = \frac{ 2^{\frac{3}{2}-\alpha}\sqrt{\pi}  u^{-\frac{1}{2}+\alpha} }{\Gamma(\alpha)}
  K_{-\frac{1}{2}+\alpha}(u)\,,
\eeq
and reduce to exponential functions for integer values (i.e., for Bessel functions of half-integer orders).

Starting at order $G^3$ (i.e, from $W^{G^3}$) new special functions arise.
It has been shown in Ref. \cite{Bini:2024ijq} (working at 2PN accuracy) that the following further families of integrals appear 
\bea
\label{all_integrals}
Q_{\alpha}^{\rm at}(u)&=&  \int dT \frac{e^{iu T}}{(1+T^2)^{\alpha}}{\rm arctan}\left(T\right)
\,,\nonumber\\
Q_{\alpha}^{\rm as}(u)&=& \int dT \frac{e^{iu T}}{(1+T^2)^{\alpha}}{\rm arcsinh}(T)
\,,\nonumber\\
Q_{\alpha}^{{\rm as}^2}(u)&=& \int dT \frac{e^{iu T}}{(1+T^2)^{\alpha}}{\rm arcsinh}^2(T)\,,
\eea
the index $\alpha$ taking half-integer values except for $Q_{\alpha}^{\rm as}$, where it takes integer values.
Using the integration by parts (IBP) identities (see Ref. \cite{Bini:2024ijq} for details) these integrals can be reduced to the three master integrals $Q^{\rm at}_{\frac12}(u)$, $Q^{\rm as}_1(u)$, $Q^{\rm as^2}_{\frac12}(u)$, and their first derivatives with respect to $u$.
Their explicit analytic expressions in terms of Meijer G functions are given in Appendix A of Ref. \cite{Bini:2024ijq}.
Starting from $O(\frac1{c^9})$ the further master integral $Q^{\rm at}_1(u)$ enters
\beq
Q_{1}^{\rm at}(u)=i\frac{\pi}{2}\left[e^{u}{\rm Ei}_1(2u)+e^{-u}\left(\gamma+\ln(2)+\ln(u)\right)\right]\,,
\eeq
where ${\rm Ei}_1(x)$ is the exponential integral, which can be in turn expressed in terms of the first derivative of the Bessel K function with respect to the order with half integer values as follows 
\beq
\frac{d}{d\nu} K_\nu(x)\bigg|_{\nu=\pm\frac12}=\pm\sqrt{\frac{\pi}{2x}}{\rm Ei}_1(2x)e^x\,.
\eeq
Note that one can define tilde quantities $Q_{\frac12}^{\rm at}(u)=i \widetilde Q_{\frac12}^{\rm at}(u)$, $Q_{1}^{\rm as}(u)=i \widetilde Q_{1}^{\rm as}(u)$, and $Q_{1}^{\rm at}(u)=i \widetilde Q_{1}^{\rm at}(u)$, which are real functions of $u$.

For increasing PM-PN order several new families of integrals appear. 
However, for each family one can derive from IBP identities recurrence relations which allow for order reduction, so that at the end only few master integrals remain.
The full structure of these integrals at higher PM accuracy will be discussed in a forthcoming paper. 

At $O(G^4)$ (i.e, in $W^{G^4}$) with 6PN accuracy there is the following further family to be considered, 
\beq
Q_{\alpha}^{\rm at\,as}(u)= \int dT \frac{e^{iu T}}{(1+T^2)^{\alpha}}{\rm arctan}(T)\,{\rm arcsinh}(T)\,,
\eeq
with the only new master integral $Q^{\rm at\,as}_{\frac12}(u)$ upon reduction given by
\beq
Q_{\frac12}^{\rm at\,as}(u)= \frac{\pi}{2}\widetilde Q_{\frac12}^{\rm at}(u)+f_{\frac12}^{\rm at\,as}(u)\,,
\eeq
with
\bea
f_{\frac12}^{\rm at\,as}(u)&=&\pi  K_0(u) \left[u \, _3F_3\left(1,1,\frac{3}{2};2,2,2;-2 u\right)\right.\nonumber\\
&&\left.
-\log (u)-\gamma +\log (2)\right]\nonumber\\
&&
-\pi ^{3/2} I_0(u) G_{2,3}^{3,0}\left(2
   u\left|
\begin{array}{c}
 \frac{1}{2},1 \\
 0,0,0 \\
\end{array}
\right.\right)\,.
\eea

{\it Results.}--
I computed the (reduced) waveform modes \eqref{Wlmdef2} through $O(G^4,\frac1{c^{12}})$, corresponding to the 5PM-6PN accuracy for the waveform.
The explicit expression for the 1SF rescaled waveform \eqref{Wfin} by taking the sum over $l,m$ (up to $l=14$) at $O(G^3)$ and $O(G^4)$, i.e., ${\mathcal W}^{G^3}$ and ${\mathcal W}^{G^4}$, is given in an ancillary file \cite{supmat} in terms of the Bessel functions $K_0(u)$ and $K_1(u)$ and their second derivatives with respect to the order as well as the master integrals $\tilde Q_{\frac12}^{\rm at}(u)$, $\tilde Q_{1}^{\rm as}(u)$, $\tilde Q_{1}^{\rm at}(u)$, and $Q_{\frac12}^{\rm at\,as}(u)$.
The combined PM-PN expansion is done as a power series expansion in both parameters $Gm_2/b$ (labeling the PM order) and $p_\infty$ (each power of which corresponding to a half-PN order).
The full angular dependence is conveniently expressed in terms of the angular variables $y=e^{i\theta}$ and $z=e^{i\phi}$. 

The current accuracy of the scattering waveform for nonspinning bodies is the one-loop level (i.e., $O(G^2)$ in the rescaled waveform) from the amplitude side, and 4PM-2PN (i.e., $O(G^3,\frac1{c^4})$) from the MPM side.
The extreme-mass-ratio limit of these results thus provides a useful check for my results at lower PM orders.
A ready-to-use expression for the complete one-loop waveform in this limit has been recently given in Ref. \cite{Heissenberg:2025fcr} up to the 10PN order.

The final form of the EFT and MPM waveforms given in Ref. \cite{Bini:2024rsy} can be directly compared with the frequency-domain 1SF waveform computed here to first order in the mass ratio, i.e., to the linear order in $\nu$. 
I find agreement in both cases modulo a phase factor $e^{i\omega\delta t}$ associated with a time shift $\delta t$ involving the various scales
\beq
{\mathcal W}=e^{i\omega\delta t}W^{\nu^1}
\equiv W^{\nu^1}+\delta W^{\nu^1}\,,
\eeq
with $W^{\nu^1}$ denoting the linear-in-$\nu$ part of the reduced waveform for either EFT or MPM, i.e., $W=\nu W^{\nu^1}+O(\nu^2)$.
In the amplitude case I find
\beq
\delta W_{\rm EFT}^{\nu^1}=i\omega\delta t_{\rm EFT}\,GW_{\rm EFT}^{\nu^1\, G^1}+O(G^3)\,,
\eeq
with 
\beq
\delta t_{\rm EFT}=2 G m_2 \, \ln\left(\frac{4Gm_2\mu_{\rm IR}e^{\gamma}}{\sqrt{e}}\right) \,,
\eeq
where $\mu_{\rm IR}$ denotes the frequency scale of the amplitude-based formalism.

In the MPM case instead I find
\bea
\delta W_{\rm MPM}^{\nu^1}&=&i\omega\delta t_{\rm MPM}\left[W_{\rm MPM}^{\nu^1\, G^1}\left(1+\frac{i\omega\delta t_{\rm MPM}}{2}\right)G\right.\nonumber\\
&&\left.
+W_{\rm MPM}^{\nu^1\, G^2}G^2\right]+O(G^4)\,,
\eea
with
\beq
\delta t_{\rm MPM}=2 G m_2 \, \ln\left(\frac{2Gm_2}{\sqrt{e}b_0}\right) \,,
\eeq
where $b_0$ is the arbitrary scale entering the tail logarithms of the MPM formalism.

Integrating over frequencies the energy flux gives the radiated energy to infinity
\beq
E_{\rm rad}^\infty=  \frac{1}{\pi^2}\int_0^{\infty} d\omega \,\omega^2\int d\Omega|{\mathcal W}(\omega,\theta,\phi)|^2\,, 
\eeq
where $d\Omega=\sin\theta d\theta d\phi$ is the standard solid angle element.
The energy loss in a two-body scattering process is usually expressed as a power series expansion in the inverse dimensionless angular momentum as follows
\beq
\frac{E^{\rm rad}}{M} =  \nu^2   \sum_{n=3}^\infty \frac{ E_{n}}{j^n}\,,
\eeq
with coefficients $E_n$ which are functions of $\gamma$ and $\nu$, and are mostly known in a PN-expanded form (i.e., as a series expansion in the small PN parameter $p_\infty$).
Their 1SF expansion is then $E_n=E_n^{\rm 1SF}(p_\infty)+O(\nu)$.

The 6PM-1SF radiated energy turns out to be 
\begin{widetext}
\bea
E_6^{\rm 1SF}&=&
\frac{4672}{45} p_\infty
+\frac{142112}{315} p_\infty^3
+\left(\frac{9344}{45}+\frac{88576 \pi^2}{675}\right) p_\infty^4
-\frac{293992}{1701}   p_\infty^5
+\left(\frac{2898 \zeta (3)}{5}+\frac{56708}{105}+\frac{1024\pi ^2}{135}\right)p_\infty^6\nonumber\\
&+&
\left(-\frac{18955264 \ln (2p_\infty)}{23625}+\frac{177152 \pi^2}{675}+\frac{36589282372}{11694375}\right)p_\infty^7 
+\left(-\frac{18189 \zeta(3)}{40}+\frac{7899371}{34020}-\frac{378890128 \pi^2}{1091475}\right)p_\infty^8\nonumber\\
&+&
\left(\frac{114102784 \ln (2 p_\infty)}{165375}+\frac{14848 \pi^2}{4725}-\frac{411613819993}{1354421250}\right)p_\infty^9\nonumber\\
&+&
\left(-\frac{37910528 \ln (2 p_\infty)}{23625}-\frac{27923 \zeta(3)}{80}+\frac{173152516 \pi^2}{945945}+\frac{253472662979}{43659000}\right)p_\infty^{10}\nonumber\\
&+&
\left(-\frac{13090527616 \ln (2p_\infty)}{22920975}-\frac{108804224 \pi^2}{1091475}-\frac{5876582290351}{84977392500}\right)p_\infty^{11} \nonumber\\
&+&
\left(\frac{240367616 \ln (2p_\infty)}{165375}-\frac{41783531 \zeta (3)}{28160}-\frac{3532707421\pi^2}{5780775}-\frac{708037817076713}{635675040000}\right)p_\infty^{12} \nonumber\\
&+&
\left(\frac{13602602608064 \ln (2p_\infty)}{16388497125}-\frac{1177068352 \pi^2}{14189175}+\frac{21026434601624568059}{22310644446090000}\right)p_\infty^{13} 
+O(p_\infty^{14})
\,.
\eea
\end{widetext}
The first few terms (up to $O(p_\infty^7)$, second line) agree with the 3PN-accurate expression of Refs. \cite{Bini:2021gat,Bini:2022enm}.
The remaining terms (from 3.5PN to 6PN, i.e., from $O(p_\infty^8)$ to $O(p_\infty^{13})$) instead are new with this work.
This result thus improves the current knowledge of the 1SF radiated energy (see Ref. \cite{Warburton:2025ymy} for a comparison between available PM, PN, 1SF, and NR results).
The 5PM-1SF coefficient $E_5^{\rm 1SF}$ has been computed very recently in Ref. \cite{Driesse:2024feo} (exact in velocity) by using the world line quantum field theory formalism, and in Ref. \cite{Geralico:2025rof} by direct integration of the time-domain energy flux in the GSF framework.

{\it Conclusions.}--
In recent years the application of quantum-amplitude-based methods to the classical two-body scattering problem has led to a significant improvement of the state-of-the-art of PM calculations, especially in the conservative sector, where the current accuracy for the potential-graviton contribution to the scattering amplitude, the radial action, and the scattering angle is 4PM, and 5PM in the 1SF approximation (see, e.g., Ref. \cite{Bern:2024adl} and references therein).
However, taking into account dissipative effects, such as radiation-reaction and recoil, complicates matters, requiring the identification of additional terms which must be added to the potential-region calculation at each loop order (see, e.g., Refs. \cite{Damour:2020tta,Herrmann:2021tct,Bjerrum-Bohr:2021vuf,Brandhuber:2021eyq}).

Therefore, it is not surprising that the amplitude-based scattering waveform is known at the one-loop level only, i.e., 3PM order.
The initial mismatch with the MPM waveform pointed out in Ref. \cite{Bini:2023fiz} was resolved by suitably modifying the EFT waveform with the inclusion of several new contributions as discussed in Refs. \cite{Georgoudis:2023eke,Georgoudis:2024pdz,Bini:2024rsy}, leading to a complete agreement up to the 2.5PN order, i.e., the level of accuracy reached in Ref. \cite{Bini:2023fiz}.
Further adjustments may be needed at higher PN orders. 

I computed here the scattering waveform up to the 5PM-6PN level in the 1SF approximation, i.e., two orders beyond the EFT waveform, even if restricted to the first order in the mass-ratio, thus providing a test-bed for future multiloop calculations.
Extending the method used here to higher PM-PN orders is computationally expensive, with an even more complicated structure of Fourier integrals.
Further results beyond the three-loop level (i.e., starting from 6PM) will be discussed in a forthcoming paper.

{\it Acknowledgments.}--
I would like to thank Thibault Damour and Donato Bini for valuable comments. 
I'm grateful to the Istituto per le Applicazioni del Calcolo ``M. Picone,'' CNR, for past support and hospitality during the development of this project.

\end{document}